\documentstyle[rotate,epsf,times]{mn}
\begin{document}

\title{Dynamical stability and evolution of the discs of Sc 
galaxies}
\author[B. Fuchs and S. von Linden]
       {B. Fuchs$^1$ and S. von Linden$^2$ \\
 $^1$ Astronomisches Rechen-Institut Heidelberg, M\"onchhofstr. 12-14, D-69120 Germany\\
 $^2$ Landessternwarte, K\"onigstuhl, D-69117 Heidelberg, Germany}

   \date{Received; accepted}
\maketitle
\begin{abstract}

We examine the local stability of galactic discs against axisymmetric
density perturbations with special attention to the different dynamics 
of the stellar and gaseous components.
In particular the discs of the Milky Way and of NGC\,6946 are studied.
The Milky Way is shown to be stable, whereas the inner parts of NGC\,6946, a typical
Sc galaxy from the Kennicutt (1989) sample, are dynamically unstable. 
The ensuing dynamical evolution of the composite disc is studied by
numerical simulations. The evolution is so fierce that the stellar disc
heats up dynamically on a short time scale to such a degree, which seems 
to contradict the morphological appearance of the galaxy. The star formation
rate required to cool the disc dynamically is estimated. Even if the star 
formation rate in NGC\,6946 is at present high enough to meet this requirement,
it is argued that the discs of Sc galaxies cannot sustain such
a high star formation rate for longer periods.

\end{abstract}


\section{Introduction}

It has been long suspected that some Sc galaxies are so gas rich that 
their gaseous discs reach the threshold of dynamical instability (Quirk 1972).
In an influential study Kennicutt (1989) has shown for a considerable
sample of 
Sc galaxies by careful analysis of the distribution of atomic and molecular
hydrogen in each galaxy that in the inner parts of the gaseous discs of the galaxies the 
Toomre (1964) stability condition is indeed violated. Kennicutt argues
further that, since the galactocentric distances of threshold of
dynamical instability coincide closely with the outer boundaries of 
the HII region discs of the galaxies, dynamical instabilities have led to the 
enhanced massive star formation rate in the inner parts of Sc galaxies.
There are, however, counter examples in his sample, notably M33 and 
NGC\,2403, which do not reach the threshold level. In addition
Ferguson et al. (1994) have shown that HII regions can be also
found in the outer -- stable -- parts of the discs of galaxies. 

Even though it is intuitively plausible that dynamical instability
leads eventually to enhanced star formation as observed in the Sc
galaxies, we
wish to point out that the existence of an unstable gas disc has grave
consequences for the dynamics not only of the gaseous disc but for the
stellar disc as well. For this purpose we examine in section 2
stability criteria for composite stellar and gas discs and derive a 
dispersion relation in order to estimate the time scale on which the 
instabilities develop. In section 3 we apply the stability criterion
to the discs of the Milky Way and NGC\,6946, a typical representative 
from the Kennicutt (1989) sample. The disc of NGC\,6946 is shown to be
dynamically unstable and we study its fierce dynamical evolution by
numerical simulations in section 4. The implications of these 
simulations are discussed in the final section.

\section{Stability criterion}

The dynamical stability of the composite gaseous and stellar disc is examined,
which gives a local stability criterion. Spiral instabilities and the
bar instability, which are less violent, are not considered in this section.
In previous stability studies of multicomponent galactic disc models the stellar disc 
has usually been described by Jeans equations (Biermann 1975, Jog \& Solomon
1984a,b, Bertin \& Romeo 1988, Elmegreen 1995, Jog 1996). Since the shape of the 
velocity distribution of the stars governs to some degree the stability of the 
stellar disc we prefer a full stellar--dynamical treatment.

It has become customary to describe the stellar velocity distribution by  a
Schwarzschild distribution. Wielen and Fuchs (1983), however, have pointed out
that a velocity distribution with an exponential shape appears to be more
realistic, at least for stars in the Milky Way, because the velocity
distribution is made up of many generations of stars with velocity dispersions
varying according to their age. First attempts to model such a distribution in
the context of stability studies have been made by Morozov (1981).

\begin{figure}
     \epsfysize=7cm
  \epsffile{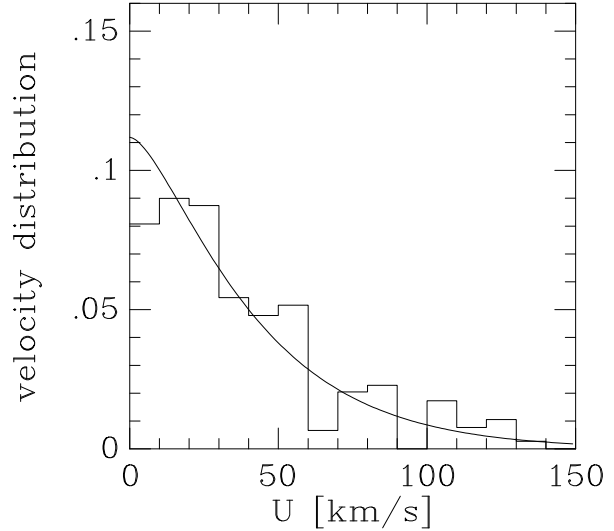}
\caption[]{\label{fig1} Distribution of -- radial -- $U$ velocities of 317 McCormick stars
in the solar neighbourhood. The observed distribution is approximated by the
projection of an exponential distribution with a radial velocity dispersion of
48 km/s.}
\end{figure}

In Fig.\,\ref{fig1} a one--dimensional projection of a distribution function of the form
\begin{equation}
f_0 = {3 \over 2 \pi \sigma_U \sigma_V} \exp - \sqrt{3} 
\sqrt{\left( {U\over \sigma_U}\right)^2 + \left( {V\over \sigma_V}\right)^2}
\end{equation}
is shown in comparison with the observed velocity distribution of the McCormick
stars -- a kinematically unbiased sample  -- in the Third Catalogue of Nearby
Stars (Gliese and Jahrei{\ss} 1991). In equation (1) $U$ and $V$ are the radial
and circumferential velocity components of the stars, respectively, and $\sigma_U$ 
and $\sigma_V$ are the corresponding second moments of the velocity
distribution which we call loosely in the following velocity dispersions.
The ratio of the velocity dispersions is given by the epicyclic ratio
${\sigma_V / \sigma_U}={-2B /\kappa}$, with $\kappa$
the epicyclic frequency and
$B$ Oort's constant.
As can be seen in Fig.\,\ref{fig1} there are a lot of low velocity
stars, which is well modelled by the distribution function (1).

In his original study of the stability of stellar discs against 
axisymmetric perturbations Toomre (1964) assumed a Schwarzschild distribution
for the stellar velocities. Graham (1965) and independently Toomre
(cited in Graham (1965)) generalised the results to arbitrary
distribution functions depending on the radial action integral
$ U^2 + {\kappa^2\over 4B^2} V^2$ (see also the appendix). The 
distribution function (1) is of this type. Graham (1965) did
not consider explicitly the case of an exponential distribution. But
Toomre has treated this case in his unpublished material,
which he kindly made available to us. The main result is that in case of an 
exponential distribution function the standard Toomre stability
parameter is slightly modified as
\begin{equation}
Q'={\kappa \sigma_U \over 3.944 G \Sigma_0}, \end{equation}
where $\Sigma_0$ denotes the surface density of the disc and $G$ is the
constant of gravitation. The numerical factor 3.944 replaces a factor
3.36 in the case of a Schwarzschild distribution or a factor $\pi$ in
the case of an ideal gas. This indicates that a stellar disc with
an exponential 
distribution is slightly more unstable than a disc with a 
Schwarzschild distribution, which has fewer low velocity stars.

We treat the interstellar gas as an isothermal gas, which will be 
sufficient to find the transition of the composite disc to
gravitational instability (Cowie 1981). Graham (1967) has
indicated how to extend his or Toomre's work to multicomponent disc models.
Unfortunately, neither he nor Toomre have treated the case of an 
isothermal gas disc inbedded in a stellar disc with an exponential velocity
distribution, which we consider here. In the appendix we discuss the stability of
such a disc against axisymmetric perturbations adapting the
analyses of Toomre and Graham to the present problem. The domain of neutrally 
stable perturbations in parameter space,
i.e. the stability parameters of the gas and stellar discs, respectively,
and the gas--to--stellar surface density ratio, indicates then which discs
are dynamically stable. For the cases, where the stability condition is
violated, we derive a dispersion relation, which allows the
determination of the wave
length and exponential rise time of the fastest growing perturbations.

\section{Stability of Galactic Discs}

The stability criterion developed in section 2 is applied to the discs of the
Milky Way and NGC\,6946.

\subsection{Milky Way}

Table 1 summarises the parameters of the Galactic disc which we adopt.
\begin{table}\centering
\caption[]{Local Parameters of the Galactic Disc}
\tabcolsep0.9cm
\begin{tabular}{rllll}\hline\hline
\rule[-4mm]{0mm}{10mm}
$\kappa$:           &            36.6      & km/s/kpc$^{a}$\\[-1.4ex]
$\sigma_U$:         &            48        & km/s$^b$\\
$\Sigma_*$ + $\Sigma_g$:       & 50        & $\cal M$$_\odot$/pc$^{{2}^{c}}$\\
$\Sigma_g$:                    & 7.5       & $\cal M$$_\odot$/pc$^{{2}^{d}}$\\
c:                             & 6         & km/s\\ \hline
\end{tabular}
\begin{tabular}{p{0.3cm}l}
$^a$    &  IAU 1984 galactic constants and assuming a\\
        &  flat rotation curve\\
$^b$    &  from solar neighbourhood data (cf.\,Fig.\,\ref{fig1})\\
$^c$    &  Kuijken and Gilmore 1989a,b,c\\
$^d$    &  Dame 1993, multiplied by a factor of 1.4 to \\
        &  account for heavy elements\\
\end{tabular}
\end{table}

The resulting stability parameters are $Q_*$ = 2.4 and $Q_g$ = 2.2,
respectively.
On the other hand, using again the parameters given in Table\,1, the 
curve of neutrally stable perturbations
derived from equation (A19) may be calculated as shown in Fig.\,2.
The
minimal
value of the stability parameter $Q_g$ required to stabilise the disc is
$Q_{g,min}$ = 1.1 implying that the Galactic disc is {\em stable} in the solar
neighbourhood.

The stability of the inner and outer parts of the Galaxy may be analysed in the
same way. The surface density and -- squared -- velocity dispersion of the
stars can be extrapolated by an exponential law with a scale length of 4.4 kpc
(Lewis \& Freeman 1989). The surface density of the interstellar gas is taken
from Dame (1993) and the turbulent velocity dispersion is assumed to be
constant. The resulting stability parameters are listed in Table 2, implying
{\em stability} of the Galactic disc against axisymmetric perturbations in the
inner and outer parts as well.
Similar conclusions were drawn by Elmegreen (1995) and Jog (1996) on the basis
of their models.

Dynamical stability of the disc is consistent with the observation
that there is no enhanced
massive star formation such as in the Sc galaxies of the Kennicutt (1989)
sample in the Milky Way. As was stressed previously by Ferguson
et al. (1994) one finds, on the other hand, still star formation
at moderate rates in
galactic discs, which are dynamically stable in the sense discussed here.
Apparently the complex structure of the interstellar medium,
which is not modelled here, still allows localised Jeans collapse. 
It appears that dynamical instability is not a necessary condition
for star formation but greatly enhances the star formation rate. 

The radial scale length of the Milky Way is not well constrained and,
as can be seen in the synopsis by Sackett (1997), may be
for the main populations of the stars in the Galactic disc
as small as 2.5 kpc. This would make the inner parts
of the Galactic stellar disc more unstable. However, the effect is not
significant. Assuming that the Galactic disc has a constant vertical scale
height as function of galactocentric radius one would expect the
-- squared -- velocity dispersion
of the stars to have the same radial gradient as the surface density
of the stars, which implies $Q_\star = 2.1$ at $R=$ 4.5 kpc.

The finite thickness of the disc, which has not
been considered here explicitly, has a further stabilising effect on the disc.

\begin{table}
\caption[]{
Stability Parameters of the Galactic Disc}
\begin{tabular}{lccc}\hline \hline
\rule[-4mm]{0mm}{10mm}
& R = 4.5 kpc
& R$_\odot$ = 8.5 kpc
& R = 12 kpc\\[2ex] \hline \\[-2ex]

$Q_*$       & 2.9    &  2.4   &  2.6\\[2ex]
$Q_g$       & 2.1    &  2.2   &  1.7\\[2ex]
$Q_{g,min}$ & 1.1    &  1.1   &  1.3\\ \hline
\end{tabular}
\end{table}

It is interesting to see whether due to the age dependence of the velocity
dispersion of the stars the Galactic disc was dynamically unstable at
earlier stages of its evolution. The velocity dispersion varies today as
$\sigma_U^2 \propto\tau$, with $\tau$ denoting
the ages of the stars (Wielen 1977). So,
assuming
a constant star formation rate, it is to be expected that at an earlier epoch
$t_{ep}$ the velocity dispersion of all  stars which were born until then
was
\begin{equation}
{\overline{\sigma_{U}^2}} = \frac{1}{t_{ep}}\,\,\,
\int\limits_0^{t_{ep}}\,\,\,
\sigma_U^2 (\tau)\,\, d \tau = \frac{1}{2}\,\, \sigma_U^2\,\, (\tau =
t{{_e}{_p}}).
\end{equation}
Table 3 shows the evolution of the stability parameters of the Galactic disc
due to this effect assuming that the other parameters of the disc were the same
as today.

\begin{table}
\caption[]{
Stability Parameters during the Evolution
of the Galactic Disc \\
(R = R$_\odot$)}
\tabcolsep0.6cm
\begin{tabular}{lccc}\hline\hline
\multicolumn{1}{c}{\rule[-4mm]{0mm}{10mm}t$_{ep}$}
& $Q_*$
& $Q_g$
& $Q_{g,min}$\\ \hline \\[-2ex]

10$^{10}$ yrs          &  2.4   & 2.2    &  1.1\\[2ex]
5$^.$10$^9$ yrs        &  1.7   & 2.2    &  1.4\\[2ex]
2.5$^.$10$^9$ yrs      &  1.2   & 2.2    &  5.5\\[2ex]
1.25$^.$10$^9$ yrs     &  0.9   & 2.2    & $>$ 20\\ \hline
\end{tabular}
\end{table}

\begin{figure}[t]
     \epsfysize=7cm
  \epsffile{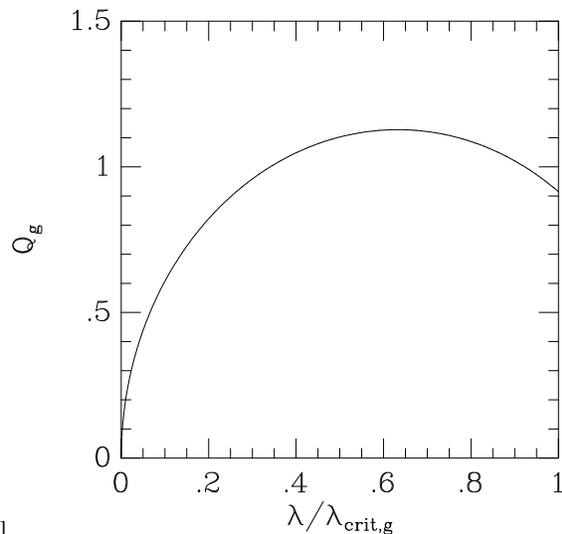}
\caption[]{\label{fig3}
Curve of neutrally stable axisymmetric perturbations of a gas
disc embedded in a stellar disc. Local parameters of the Galactic disc
have been used.}
\end{figure}

Apparently, this indicates instability of the disc during early stages of
galactic evolution.
This would be in line with Kennicutt et al.'s (1994) argument that early
type spirals had apparently an initial burst of star formation.
However, we believe, for reasons explained in the next sections, phases of
dynamical instability extending over more than $10^9$ years unlikely.
Probably, the surface
density of the disc was then lower than today and a lot of
material might have been accreted later onto the disc (Gunn 1982). If, for
example,
the surface density was only one half of its present value and the gas density
equal to the stellar density, the stability parameters were $Q_*$ = 4, $Q_g$ =
1.3, whereas the minimal value of the parameter required to stabilise the disc
was $Q_{g,min}$ = 1.2.

\subsection{NGC 6946}

NGC\,6946 is a typical representative of the Sc galaxies studied by Kennicutt
(1989). The absolute mass as well as the radial distribution of atomic and
molecular hydrogen in NGC\,6946 have been determined by Tacconi \& Young
(1986). The rotation curve has been observed by Carignan et al.\,(1990), who
have also constructed a mass model comprising a disc and dark halo component.
Unfortunately, the stellar velocity dispersion is not known, but one may derive
an estimate from the vertical scale height of the disc. Friese et al.\,(1995)
have shown in a statistical flattening analysis of faint spiral galaxies in the
ESO--Uppsala catalogue that the typical intrinsic ratio of vertical to radial
scale lengths of Scd galaxies is of the order of $z_0/h$ = 0.25 implying $z_0$
= 1.4 kpc. This may be used
with the vertical hydrostatic equilibrium condition to derive an estimate of
the vertical velocity dispersion of the stars,

\begin{equation}
{\sigma}^2_W = {\cal F} \pi G \Sigma_0 z_0 .
\end{equation}

$\cal F$ is a form factor which is equal to 1 for an isothermal disc with a
vertical $sech^2$ density profile and about 2 for the Milky Way disc (Friese et
al.\,1995).
The same value is also adopted for the disc of NGC\,6946. The radial velocity
dispersion
is estimated by using the same axial ratio of velocity ellipsoid as in the
Milky Way, $\sigma_U^2/\sigma_{W}^2 = 1 + \kappa^2/4B^2$. In
columns 5, 8, and 9 of Tab. 4 the radial variations of the stability
parameters,
which have been calculated including finite thickness corrections, are shown.
We note that the stability parameter of the stellar disc determined in this way, $Q_\star \approx 2$, is in good agreement with values determined by
Bottema (1993) for galaxies for which kinematic data are available.

\begin{table*}\centering
\caption[]{Radial Variation of Stability Parameters
of the Disc of NGC\,6946 (d = 10 Mpc)}
\tabcolsep0.5cm
\begin{tabular}{ccccccccccc}\hline\hline
\rule[-4mm]{0mm}{10mm}
R  & $\Sigma_*$ &  $\kappa$ & $\sigma_U$ & $Q_*$ & $\Sigma_g^a$& $c$ &
 $Q_g$ & $Q^b_{g,min}$ & -$\Im$($\omega$)/$\kappa$
& $\lambda_{max}$ \\ \hline \\[-2ex]
5   & 133  &  50  &  114  &  2.5   & 38 & 6 & 0.6 & 0.9 & 0.9 & 0.7 \\
10  &  55  &  23  &   89  &  2.2   & 21 & 6 & 0.5 & 1.1 & 1.4 & 1.0 \\
12  &  38  &  20  &   76  &  2.3   & 17 & 6 & 0.5 & 1.1 & 1.4 & 1.2 \\
14  &  27  &  17  &   66  &  2.4   & 14 & 6 & 0.5 & 1.1 & 1.4 & 1.5 \\
\hline \\[-2ex]

16  &  19  &  15  &   52  &  2.3   &  6 & 6 & 1.0 & 1.0 & --- & --- \\
\hline \\[-2ex]
18  &  13  &  13  &   44  &  2.6   &  5 & 6 & 1.1 & 1.0 & --- & --- \\
20  &  9   &  12  &   39  &  3.0   &  5 & 6 & 1.0 & 1.0 & --- & --- \\
$kpc$ & $\frac{M_\odot}{pc^2}$ & $\frac{km/s}{kpc}$ & $km/s$ &  &
$\frac{M_\odot}{pc^2}$ & $km/s$ &  &  &  & $kpc$ \\ \hline

\end{tabular}
\begin{tabular}{p{0.3cm}l}
$^a$    & Molecular hydrogen surface densities have been rescaled to the CO
  density -- to -- H$_2$ column density conversion factor given by \\
 & Dame (1993). Atomic and molecular hydrogen surface densities have been
   multiplied by a factor 1.4 to account for heavy elements.\\
$^b$  & Corrected for the finite thickness of the discs, assuming $z_{0,\star}=1.4$ kpc
 and $z_{0,g}=50$ pc.\\
\end{tabular}
\end{table*}

\begin{figure}
     \epsfysize=7cm
  \epsffile{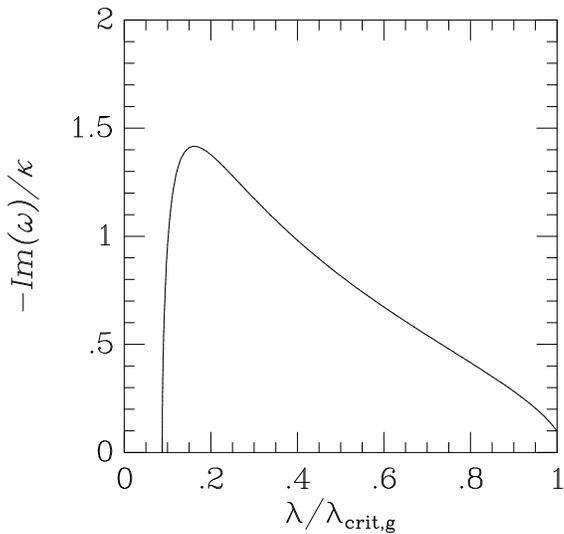}
\caption[]{\label{fig4}Dispersion relation of exponentially growing axisymmetric
perturbations of a composite gas and stellar disc. Parameters of the disc of
NGC 6946 at galactocentric distance R = 10 kpc have been used.}
\end{figure}

There is a distinct drop of the stability parameter of the gas disc $Q_g$
inside R = 16 kpc, which, as was pointed out by Kennicutt (1989), is correlated
with the outer boundary of the HII region disc of NGC\,6946. As can be seen
from
column 9 of Table\,4 the stability criterion as presented here indicates
in  accordance with Kennicutt's conclusions that the inner parts of the disc of
NGC\,6946 are actually {\em dynamically unstable}.
This is mainly due to the large gas content of NGC\,6949, which is quite
typical for late type spiral galaxies (Kennicutt et al. 1994).
The exponential growth rates
as well as the most unstable wavelengths, which may be calculated from the
combined
dispersion relation (A19)
are illustrated in Fig.\,\ref{fig4} and are given in columns
10 and
11 of Table\, 4. The e-folding rise time of the instability is rather short,
only 3$\cdot$10$^7$ yrs at R = 10 kpc corresponding to 0.1 epicyclic
periods. The
unstable wavelengths are short compared to the radial extent of the galactic
disc.

Using equations (\ref{a12}) and (\ref{aend}) one can show that, even if the
instability grows at the same rate in the gaseous and the stellar discs,
the amplitude of the disturbance of the stellar disc is much smaller than in
the gaseous disc. For the parameters adopted in Table\,4 the relative
star--to--gas density contrast is typically of the order of 10$^{-3}$ so that,
even when the perturbations of the gas disc become highly non--linear, the
stellar disc is only mildly perturbed.

\section{Numerical simulations}
\begin{figure}
     \epsfysize=6cm
  \rotate[r]{\epsffile[50 60 305 400]{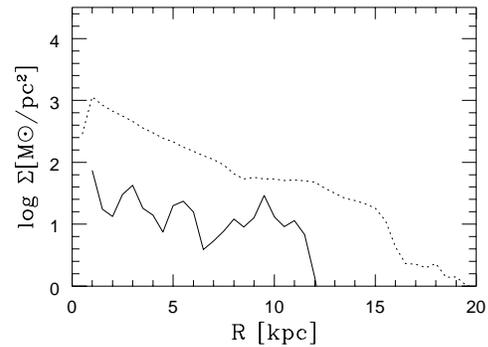}}
\caption[]{\label{fig5a} Surface density distribution of the stellar
(dotted line) and the gas (solid line) at time step t=1.2
$\cdot 10^7$ yrs of the simulation.}
\end{figure}
\begin{figure}
     \epsfysize=6cm
  \rotate[r]{\epsffile[50 60 305 400]{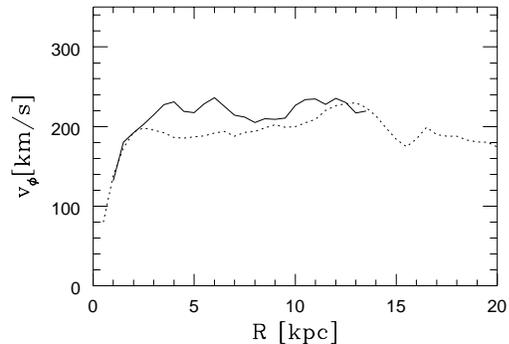}}
\caption[]{\label{fig5b} Rotation curve of the stars (dotted line)
and the gas (solid line) at time step
t=$1.2\cdot 10^7$ yrs of the simulation.}
\end{figure}

The onset of the instability into the non--linear regime  can be followed by
numerical simulations.
Simulations of a dynamically unstable gas disc inbedded in a stellar
disc were first carried out by Carlberg \& Freedman (1985). We have
rerun similar simulations in order to analyse them in the present
context. For this purpose we have used a code developed by
F.~Combes and collaborators (cf.~ Casoli and Combes 1982, Combes and
G\'{e}rin 1985 for the full details). The code implements a two--dimensional
stellar disc (N$_{star}= 38\,000)$
into which interstellar gas clouds are embedded
(N$_{clouds}\leq 38\,000$). This composite
disc is surrounded by a rigid bulge and dark halo potential. The
gravitational potential of the disc is calculated by a standard
particle--mesh scheme and the orbits of the stars and the clouds are integrated
numerically. The stars 
interact only by -- softened -- gravitational forces,
while the gas clouds 
may interact inelastically. This is simulated in an
elaborate cloud--in--cell scheme, which describes the coalescence and
fragmentation of the gas clouds. In this way a mass spectrum of the gas clouds
is established.
The various parameters of the collision scheme have been adjusted by
Combes \& G\'{e}rin (1985) in such a way that the mass spectrum of the gas
clouds resembles the mass spectrum of molecular cloud complexes in
the Milky Way. A finite lifetime of 4$\cdot$10$^7$ years due to massive
star formation is assumed for the giant molecular clouds (GMC) at the
high end of the mass spectrum, $M_{\rm GMC} >
2 \cdot 10^5 {\rm M}_\odot$.
After that the clouds are disrupted
into small fragments expanding initially isotropically at relative speeds of
10 km/s.

The softening length of the gravitational forces is 500\,pc.
Following Romeo (1994) we have adjusted it in this way, because on one hand
it is still considerably smaller than the critical wavelengths of the
interstellar gas and stellar discs, respectively (see below). On the other
hand, it is of the same order of magnitude of the expected vertical scale
height of the Galactic disc, so that the stabilising effect of a finite
thickness of the disc is taken into account to some extent in our
two-dimensional simulations. 

\begin{figure}
     \epsfysize=8.0cm
  \rotate[r]{\epsffile[170 160 490 590]{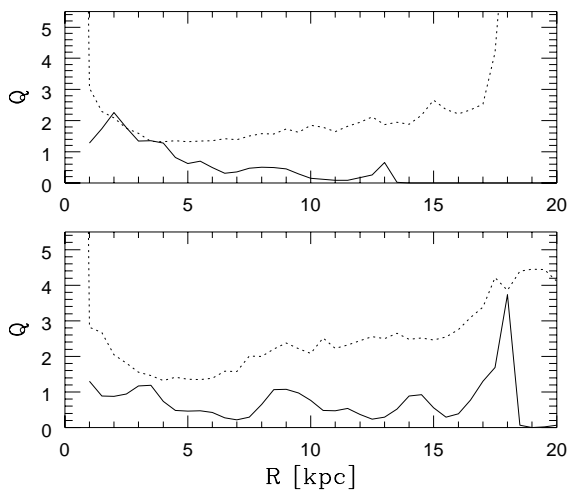}}
\caption[]{\label{fig5c} Stability parameters of the stellar disc
(dotted lines) and the gas disc (solid lines) at time steps
t=6$\cdot 10^7$ yrs (upper panel) and t=$18\cdot 10^7$ yrs (lower
panel) of the simulation.}
\end{figure}

\begin{table}\centering
\caption[]{\label{simtab}Parameter adopted for the simulation}
\tabcolsep0.6cm
\begin{tabular}{lcl}\hline \hline
stellar disc mass      $M_d$  & 7.7 &$10^{10} {\rm M}_\odot$\\
scale length $d$ &4 &kpc\\
truncation radius & 32 & kpc\\[2ex]

gas disc mass  $M_g$   & 0.6 &$10^{10} {\rm M}_\odot$\\
exponential scale length  & 16 &kpc\\[2ex]

dark--halo mass $M_h$ & 10 &$10^{10} {\rm M}_\odot$  \\
core radius    & 24 &kpc\\[2ex]

bulge mass     $M_b$ & 2.4 &$10^{10} {\rm M}_\odot$\\
scale length    &1.5 &kpc\\[2ex]

mass ratio $(M_d+M_g)/M_{\rm tot}$ & 0.40&\\
\hline

\end{tabular}
\end{table}

The stars are initially distributed according to a Toomre disc (Toomre 1963),
\begin{equation}
\Sigma_d(R)={M_d \cdot d \over 2 \pi} \cdot (R^2 + d^2 )^{-3/2}\, ,
\end{equation}
where $M_d$ denotes the mass of the disc and $d$ is the radial scale length.
The gas clouds are distributed in an exponential disc.
The halo and the bulge are modelled by Plummer spheres. 
The parameters which we have adopted for the various components
of the model are summarised in Table\,\ref{simtab}. 
The begin of the simulation is illustrated in Figs.~\ref{fig5a}
to \ref{fig5c}, where
the surface densities, the rotation curve, and the stability parameters
are shown.
Fig.~\ref{fig5c} shows that the 
the combined disc of stars and interstellar gas clouds is {\em
dynamically unstable} and resembles in that the inner parts of the
Sc galaxies in the sample of Kennicutt (1989).
The next steps of the evolution of the disc are
shown in Fig.~\ref{fig6} at multiple intervals of $\Delta$t = 1.2$\cdot$10$^7$ years which
correspond to about 0.1 epicyclic periods. As expected ring--like density
perturbations appear immediately in the gas disc. The wave lengths of these
perturbations are of the order  of the critical wave length of the
gas disc, $\lambda_{crit,g}$ about 2 kpc at R\,=\,10\,kpc.
The rings fragment into lumps with masses in the
range of 10$^4$ to 10$^7$ M$_\odot$. These agglomerates are so heavy that
the stellar disc responds to them by induced, `swing--amplified' spiral
structures. According to the `X=2' criterion (Toomre 1981) swing--amplification
is most effective for spirals with a number of
\begin{figure*}
     \epsfysize=22.0cm
  \epsffile[25 45 587 734]{sim100.ps}
    \caption[]{\label{fig6}Dynamical evolution of the stellar and
gaseous discs. On the left of each panel 19\,000 out of 38\,000 stars and on the
right hand side of each panel19\,000 of about 38\,000 interstellar
gas clouds are plotted at consecutive time
intervals. The time is indicated on the left in units of $10^7$ years.
The spatial size is indicated at the bottom by a bar of 10 kpc length.}
\end{figure*}

\begin{equation}
m = {1\over 2} \cdot \frac{2 \pi R}{\lambda_{crit}}
\end{equation}

spiral arms. Since $\lambda_{crit}$ = 10 kpc, the expected number of
spiral arms
is about $m$ = 3, which is typically seen in our simulations
(see for instance time step $14.4 \cdot 10^7$). The potential
troughs of the stellar disc, on the other hand, begin to trap much of the
interstellar gas and during their further evolution both the stellar disc
as well as the gas disc undergo rather synchronous, repetitive cycles of swing
amplified spiral perturbations.

After 5$\cdot$10$^8$ years the stellar disc gets heated up  dynamically
by the spiral activity so much that hardly any non--axisymmetric structure is
longer possible in the disc.

In the gaseous disc, however, there is still a lot of spiral activity. Since
the stellar disc has become inactive the critical wavelengths of the spiral
structures are much smaller and the number of expected spiral has risen to about 20 in accordance with flocculent appearance of the disc.
The same phenomenon was described by Carlberg \& Freedman (1985)

\section{Discussion}

The numerical simulations show that the discs of Sc galaxies like
NGC\,6946 are in a highly peculiar dynamical state.
The reaction of the stellar 
discs to dynamically unstable gas discs is so fierce
that they become dynamically
hot within less than $10^9$ years. On the other hand, Toomre
(1990) has argued emphatically that Sc galaxies must have stellar discs, which
are dynamically active, because otherwise the morphological appearance
would be quite different from what is actually
observed. This can be clearly seen in Fig.\,7, when one compares the frames 
corresponding of say $1.44\cdot 10^8$ years and 4.8$\cdot 10^8$ years
with an optical image of the galaxy (Sandage \& Bedke, 1988). Thus the stellar
discs in Sc galaxies must be effectively cooled dynamically by newly formed stars on
low velocity dispersion orbits. The star formation rate required to keep the stellar
discs in a steady state can be roughly estimated as follows.
After $5\cdot 10^8$ years the stellar disc has heated up so much
that it tunes out of spiral activity and the -- squared -- stability parameter has risen to
\begin{equation}
Q_\star^2= {\kappa^2 \sigma_\star^2\over (3.36 G \Sigma_\star)^2}\approx 6.
\label{6a}\end{equation}
Newly formed stars will lower this to
\begin{equation}
Q_\star^{\prime 2}= {\kappa^2 (\Sigma_\star \sigma_\star^2 + \delta_\star
\Sigma_\star \sigma_{\star, i}^2)\over (3.36 G)^2
(1+\delta_\star)^3
\Sigma_\star^3},\label{6b}\end{equation}
where we assume that the
surface density of the stellar disc has risen due to star formation to $\Sigma_\star^\prime=\Sigma_\star(1+\delta_\star)$
and that the  -- squared -- velocity dispersion of the mixture of older and younger stars can be
approximately estimated by an mass weighted average of the velocity
dispersions of the components. Combining equation (\ref{6a}) and (\ref{6b})
gives
\begin{equation}
(1+\delta_\star)^3={Q_\star^2\over Q_\star^{\prime 2}} (1+\delta_\star
{{\sigma_{\star, i}}^2\over \sigma_\star^2} )\simeq
{Q_\star^2\over Q_\star^{\prime 2}},\end{equation}
if the velocity dispersion of the  newly born
stars, $\sigma_{\star, i}$,
is much smaller than the average velocity dispersion of the stars.
In our simulations the stability parameter $Q_\star$
rose from about initially 2 to 2.5 at $t=4.8 \cdot 10^5$ yrs.
According to equation (9) about 40 \%
of the mass of the stellar disc is required per $10^9$
years in the form of newly born stars to keep the stellar disc in a steady
dynamical state. Obviously this material must be provided by the gaseous
disc. It is interesting to compare this estimate of the  gas consumption
rate with the actual gas consumption rate in NGC\,6946.
This  can be deduced from the star formation rate which in turn can be
quantitatively estimated from the H$_\alpha$ flux using the relation of
Kennicutt (1983). In Table 6 we give the radial distribution of the
star formation rate determined from the extinction corrected H$_\alpha$ surface
emissivity of NGC\,6946 (Devereux \& Young 1993). The surface densities of the
gaseous and stellar discs shown in Table 6 are taken form Table 4.
In the last column
of Table 6 we give the relative increase of the stellar surface density
per $10^9$ years, $\delta_\star$, estimated from the previous parameters.
\begin{table}\centering
\caption[]{Radial variation of the stellar disc mass increase per
$10^9$ yrs in NGC\,6946}
\tabcolsep0.4cm
\begin{tabular}{cccccc}\hline\hline
\rule[-4mm]{0mm}{10mm}
R  & $SFR$ &  $\Sigma_\star$ & $\Sigma_{gas}$ & $\delta_\star$ \\ \hline \\
0   & 285  &  325  &  306  &  0.9\\
1.1  &  180  &  267  &   143  &  0.7 \\
1.2  &  71  &  219  &   95  &  0.3 \\  
4.4  &  57  &  148  &   53  &  0.4  \\ 
6.6  &  28  &  100  &   36  &  0.3 \\
8.8  &  11  &  67  &   27  &  0.2 \\
11  &  9  & 46  & 15  &  0.2 \\
13.2 & 4  & 31  &   11  &  0.1 \\
$kpc$ & $\frac{M_\odot}{pc^2}/Gyr$ & ${M_{\odot}\over pc^2}$ & ${M_{\odot}\over pc^2}$ & & \\ \hline
\end{tabular}
\end{table}
As can be seen from Table 6 the actually observed star
formation rate and thus the disc mass increase is at
present, if compared with the theoretical
estimate found above, high enough to keep the stellar disc dynamically cool.
In this aspect the disc seems to be self regulating. However, the gas
consumption rate is very large. As is shown in Table 6 star
formation seems at present to consume nearly the entire gas disc
within $10^9$ yrs. This implies either a very high gas accretion rate,
which at this magnitude seems unlikely
to us, or that the discs of the Sc galaxies of the NGC\,6946 type will
soon switch over to more quiescent dynamical states like in M33 or NGC\,2403.
In these galaxies
the gaseous discs do not reach the threshold of dynamical instability. But
even if the stability parameter $Q_g$ is slightly larger than
1, this would mean still quite a lot of spiral activity in the disc.
This might account for the still comparatively high star formation rates
observed in these galaxies (Kennicutt et al. 1994), but leads also
to considerable dynamical heating (Carlberg \& Sellwood 1984).
The same comment applies to the low surface brightness galaxies discussed by
Mihos et al. (1997).

A further interesting example is the Sb galaxy NGC\,7331.
Its radial distribution of atomic and molecular hydrogen is given by Young \&
Scoville (1982). After rescaling the molecular hydrogen densities to a modern
CO intensity--to--H$_2$ column density ratio (Dame 1993) it can be
shown that the gaseous disc of NGC\,7331 does not reach the threshold of dynamical
instability, $Q_g \tilde{>} $1. In accordance with this result the star
formation
rate seems to be in NGC\,7331 much lower than in NGC\,6946. Applying again
the Kennicutt
(1983) relation to the observed integral H$_\alpha$ flux
(Young et al. 1996) after correcting
this for internal extinction in the way suggested by Devereux \& Young (1993)
we estimate a global star formation rate of 7 M$_\odot$/yr. This is about
half the value found for NGC\,6946. Furthermore, the gaseous and, in
particular, the
stellar disc (Broeils 1995) are more massive than in NGC\,6946 so that
the relative
disc mass increase is only moderate.

Finally we note that the adopted distances to the galaxies affect the dynamical stability condition
of their gas discs (cf. Zasov \& Bizyaev 1996).
The surface densities and amplitudes of the rotation curves are distance
independent. The epicyclic frequency scales inverse proportional to the
distance, while
the estimate of the turbulent velocity dispersion of the gas used throughout this
study is distance independent.
Thus assuming half of the presently adopted distance to NGC\,6946, as suggested
by de Vaucouleurs (1979),
raises the stability parameter
of the gas disc close to the threshold of dynamical instability.
\section{Conclusions}

We have formulated a local stability criterion of Galactic discs against
axisymmetric density disturbances modelling the different dynamics of
the stellar
and gaseous components. The disc of the Milky Way is shown to be dynamically
stable at all Galactic radii and probably over most of its past history.
The inner parts of the disc
of NGC\,6949, a typical Sc galaxy from the Kennicutt (1989) sample,
are found to be dynamically unstable. We
have followed the ensuing
dynamical evolution of the disc by numerical simulations.
These show that such unstable discs evolve very rapidly. In order to stay in
its present state the stellar disc would have to be effectively cooled by
star formation. This seems to be actually observed in NGC\,6946.
However, the gas disc would have to be replenished by heavy accretion of gas,
amounting to several times
the present day gas disc mass during a Hubble time.

\section*{Acknowledgements}

We are grateful to Alar Toomre for his advice
and Francoise Combes for letting use us her code. The numerical simulations
were run on the YMP Cray of the HLRZ, KFA--Forschungszentrum J\"ulich.
SvL was supported by the Deutsche Forschungsgemeinschaft (SFB 328).

\appendix
\onecolumn
\section{}
The stability of the self-gravitating composite gas and stellar disc is
examined. Both disc components are approximated as infinitesimal thin
sheets.
In order to test the stability of the disc it is subjected to density
perturbations of the form
\begin{equation}
\exp i(\omega t + kR)\,,
\end{equation}
where $\omega$ and $k$ denote the --complex-- frequency and radial wave number,
respectively. Since the most unstable wavelengths $\lambda = 2\pi/k$ turn
out to be small compared to the radial extent of the disc it is sufficient to
study the stability of the disc in a localised theory (Toomre 1964).
Considering now a strip of the galactic disc, radial variations of the surface
density and of the velocity distribution of the stars and the interstellar
gas can be neglected.
\subsection{Stellar disc}

In order to describe the dynamics of the stars in the circular strip around the
galactic centre we consider the collision-less
Boltzmann equation,

\begin{equation}
\frac{\partial f}{\partial t} + [f,H] = 0,
\end{equation}

where $f$ denotes the distribution function of the stars in phase space and
$H$ is the Hamiltonian. The distribution function of the unperturbed
disc, $f_0$, is chosen according to equation (1) and normalised to the surface
density $\Sigma_0$.
Radial variations of $f_0$ are neglected. 
The -- axisymmetric -- disc response $f_1$ to a small axisymmetric
potential perturbation, $\delta \Phi = \Phi_k
\exp{ i k R}$, is calculated from
the linearised Boltzmann equation

\begin{equation}
\frac{\partial f_1}{\partial t} + [f_1,H_0] +[f_0,\delta \Phi] = 0.
\end{equation}

\noindent In the following the plane stellar orbits are described according to the
epicyclic approximation. The Hamiltonian $H_0$ is then given by

\begin{equation}
H_0 = \frac{1}{2} {\dot{R}}^2 + \frac{1}{2} R^2_0 ( {\dot{\theta}} - \Omega_0
)^2 - 2 A \Omega_0 ( R - R_0 )^2,
\qquad \mbox{or alternatively}\qquad
H_0 = \frac{J_1}{2} + \frac{A}{2B} ( J_2 - \Omega_0 R_0 )^2-{1\over2}
\Omega_0^2 R_0^2,
\end{equation}

where $A$ and $B$ denote Oort's constants. This leads to the equations of
motion

\begin{equation}
R - R_0  =  \frac{J_2 - \Omega_0 R_0}{- 2 B} + \sqrt{\frac{2 J_1}{\kappa}}
\sin{w_1},\qquad R_0 (\theta - \Omega_0 t)  =  w_2 - \frac{\sqrt{2 \kappa J_1}}{2 B}
\cos{w_1},
\end{equation}
\begin{equation}
U  =\dot{R} = \sqrt{2 \kappa J_1} \cos{w_1}, \qquad\quad
V  = R_0(\dot{\theta}-\Omega_0)+2A(R-R_0)
   = 2 B \sqrt{\frac{2 J_1}{\kappa}} \sin{w_1}, \nonumber
\end{equation}

where $R, \theta$ denote polar coordinates. $R_0$ is the mean galactocentric
radius of the strip of the disc under consideration. $\Omega_0$ is the mean
angular velocity of the stars in the strip around the galactic centre.
$J_1$ and $J_2$ are
integrals of motion, $J_1 = (U^2+\frac{\kappa^2}{4B^2}V^2)^{1/2}$ the radial
action integral and $ R_0\left( J_2-\Omega_0 R_0\right)$
the angular momentum of a star referred
to the mean radius of the strip. $w_1$ and $w_2$ denote the conjugate variables
derived from the Hamiltonian $H_0$,

\begin{equation}
w_1 = \kappa t ,\qquad \quad w_2 = \frac{A}{B} (J_2 - \Omega_0 R_0) t .
\end{equation}

The Poisson brackets in equation (A3) are then given by

\begin{equation}
[f_1,H_0] = \kappa \frac{\partial f_1}{\partial w_1} ,
\qquad \quad
[f_0,\delta \Phi] = - \frac{\partial f_0}{\partial J_1} \cdot
  \frac{\partial \delta \Phi}{\partial w_1} = \frac{\sqrt{3}}{\sigma_U}
  \delta \Phi i k \cos{w_1} \label{a7}.
\end{equation}

Since the disc response is axisymmetric no $\partial f_1 /\partial w_2$
derivatives appear in equation (\ref{a7}). This leads to the final form of the
linearised Boltzmann equation

\begin{equation}
\frac{\partial f_1}{\partial t} + \kappa \frac{\partial f_1}{\partial w_1} 
= -\Sigma_0 \frac{\kappa}{-2B} \frac{3\sqrt{3}}{2\pi\sigma_U^3}
\exp{-\left(\frac{\sqrt{3}}{\sigma_U}\sqrt{2\kappa J_1}\right)} i k
  \Phi_k\cos{w_1}\exp{i k \left(\frac{J_2 - \Omega_0
R_0}{-2B}+ \sqrt{\frac{2 J_1}{\kappa}}\sin{w_1}\right)} \label{a9}
.
\end{equation}

Assuming a time dependence according to equation (A1),

\begin{equation}
f_1 , \Phi_k \propto \exp{i \omega t},
\end{equation}

equation (\ref{a9}) can solved by combining solutions
of the homogeneous part of
the equation and a particular solution of the inhomogeneous equation found by
`variation of the constant', giving finally

\begin{equation}
f_1  =  - \frac{f_0}{\sqrt{2\kappa J_1}} e^{i \omega t}
\frac{\sqrt{3}}{\sigma_U} \Phi_k e^{i k (R - R_0)} \cdot \left\{ 1 -
\frac{\omega/\kappa}{2\sin{(\pi\omega/\kappa)}}\int_{-\pi}^{\pi}\,\,dw'_1
\exp{i \left[ \frac{\omega}{\kappa}w'_1  -
\sqrt{\frac{2J_1}{\kappa}}k\left(\sin{w_1}
+ \sin{(w_1+w'_1)}\right) \right] }\right\}.\label{a11} 
\end{equation}

In order to have a distribution function $f_1$, which is uniquely defined in
velocity space,
the integration constant of the solution (\ref{a11}) has been
chosen in such a way that $f_1$ is periodic with
respect to the angular variable $w_1$ (cf.\,Toomre 1964). The density response
of the disc to the potential perturbation is found by integrating the
distribution function $f_1$ over velocity space,

\begin{eqnarray}
\Sigma_{1,k} & = & - \int_0^{\infty}dJ_1 \int_0^{2\pi}dw_1 \Sigma_0
\frac{3\sqrt{3}\kappa}{2\pi\sigma_U^3}\exp{-
\left(\frac{\sqrt{3}}{\sigma_U}\sqrt{2\kappa J_1}\right)} 
\frac{1}{\sqrt{2\kappa J_1}}\Phi_k e^{i k (R-R_0)}
\Biggl\{  1 -
\frac{\omega/\kappa}{2\sin{(\pi\omega/\kappa)}} \nonumber
\\ & & \left. \cdot \int_{-\pi}^{\pi} dw'_1
\exp{i \left[ \frac{\omega}{\kappa}w'_1 - 
\sqrt{\frac{2J_1}{\kappa}}k\left(\sin{w_1}
+ \sin{(w_1+w'_1)}\right) \right] } \right\}.  \label{a12}
\end{eqnarray}

The surface density perturbations are assumed to be self consistent. Therefore
the density and potential perturbations have to satisfy the Poisson equation, 
\begin{equation}
\Delta\delta\Phi = 4\pi G\Sigma_1 \delta(z),
\label{lapl}\end{equation}
where $\delta(z)$ denotes a
delta function with respect to the vertical spatial coordinate $z$.
If the wave numbers $|k|$ are sufficiently large, $|k|R\gg 1$,
the Laplace operator in equation (\ref{lapl}) has nearly cartesian form
and one obtains

\begin{equation}
\delta \Phi_k \propto e^{i k (R-R_0)-k|z|} \qquad {\mbox{and}} \qquad
\Phi_{k} e^{i k (R-R_0)} = -{\Sigma_{1,k} \over  2 \pi G |k|}.
\label{a14}\end{equation}

\begin{figure}
     \epsfysize=7cm
  \epsffile{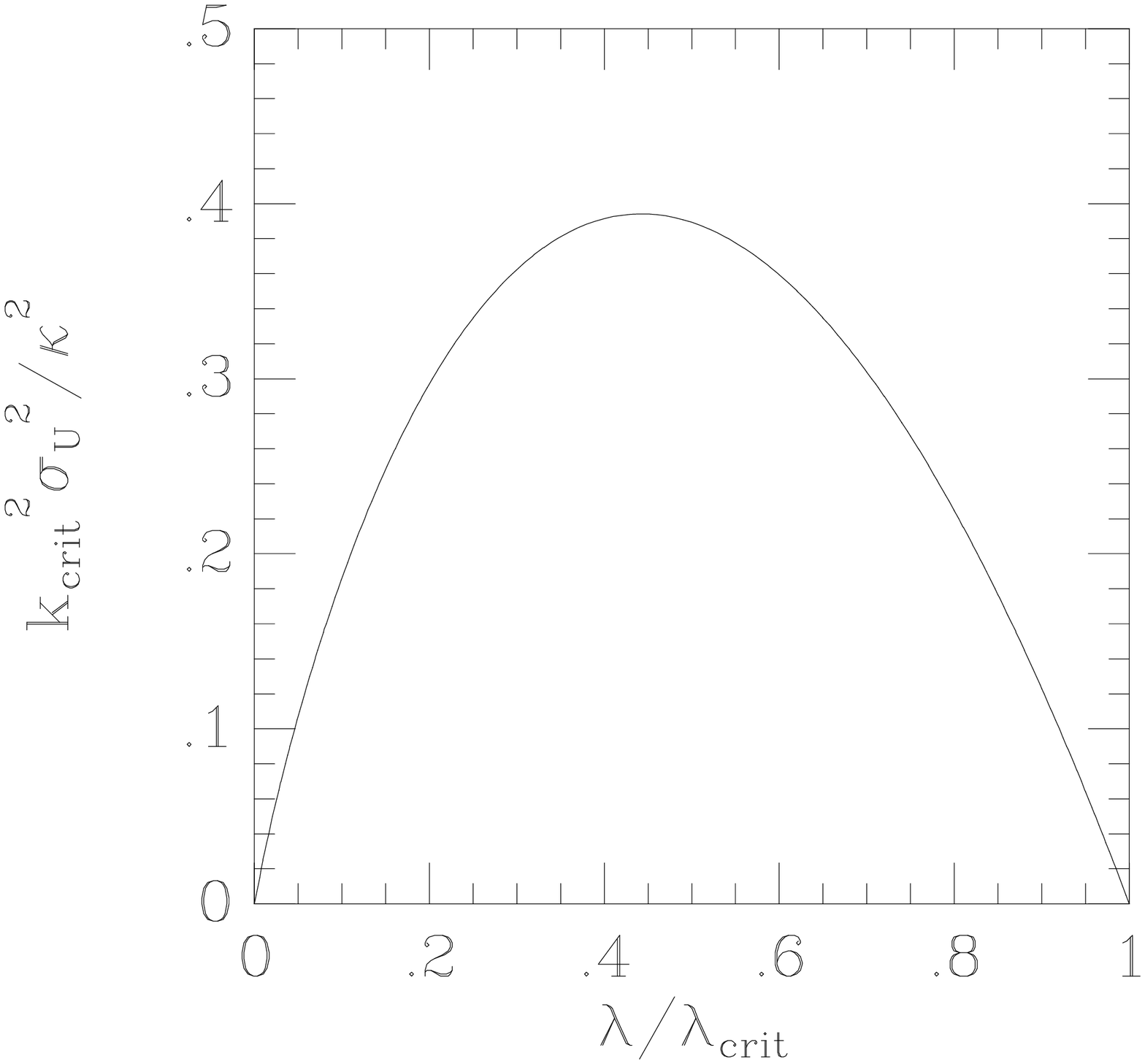}
\caption[]{\label{fig2} Curve of neutrally stable axisymmetric perturbations of a
stellar disc with an exponential velocity distribution. Unstable solutions lie
below the curve.}
\end{figure}

Thus, the amplitude of the density perturbations, $\Sigma_{1,k}$, cancels out
of equation (\ref{a12}). Finally one obtains after evaluating the quadratures with
respect to $J_1$ and $w_1$ the dispersion relation.

\begin{eqnarray}
1  =  \frac{\mid\lambda\mid}{\lambda_{\rm crit}} \frac{3}{\xi} 
         \Biggl\{ 1 
- \frac{\omega/\kappa}{2\sin(\pi\omega/\kappa)} 
    \int\limits_{-\pi}^{+\pi}
dw'_1 \frac{\exp(i{\frac{\omega}{\kappa}w'_1)}}
{\sqrt{1+\frac{4}{3}
\xi (\frac{\lambda_{\rm crit}}{\lambda})^2 \cos^2
({\frac{w'_1}{2}})}}\Biggr\}\,,
\end{eqnarray}

where $\lambda_{\rm crit}$ = 4$\pi^2G\Sigma_0/\kappa^2$ is the critical
wavelength and $\xi$ = 4$\pi^2\sigma_U^2/\lambda_{\rm crit}^2\kappa^2$.
For $\omega$ = 0 equation (A15) gives the curve of neutrally stable perturbations
in the
($\lambda$, $\xi$) parameter space as shown in Fig.\,\ref{fig2}. Exponentially unstable
perturbations lie below the curve, whereas oscillatory solutions lie above. As
can be seen from Fig.\,\ref{fig2} discs with $\xi \ge$ 0.394 are stable at all
wavelengths. This corresponds to the Toomre parameter given equation (2).

In case of a three--dimensional disc of finite thickness the vertical oscillations
of the stars have to be taken into account. In the epicyclic approximation,
however, the vertical oscillations separate from the planar motions and
the energy associated with
the vertical oscillations has simply to be added to the Hamiltonian (A4).
Vandervoort (1970a,b) has developed methods
to solve the Boltzmann and Poisson equations for such a system. A rough 
estimate of the finite thickness  effect on the stability of the disc can
be found, however, according to 
Toomre (1964). It can be shown that potential perturbations of a disc
with an effective scale height $z_0$ are reduced by a factor of
\begin{equation}
\frac{1-\exp(-k z_0)}{k z_0}
\end{equation}
with respect to the potential perturbations of an infinitesimal thin disc with
the same surface density. In order to correct the dispersion
relation for this effect, the right hand side of equation (A15) has to be
multiplied by the reduction factor (A16).

\subsection{Gaseous disc}
The gaseous component of the galactic disc is modelled by an isothermal gas.
As is well known (Binney \& Tremaine 1987), the density response of the
gaseous disc is given by

\begin{equation}
\Sigma_{g 1,k} = \Sigma_{g 0} \Phi_k \frac{k^2 e^{i k (R - R_0)}}{\omega^2
-c^2 k^2 + 4 \Omega_0 B}, \label{aend}
\end{equation}

where $\Sigma_{g 0}$ denotes the unperturbed surface density of the gaseous
disc and c is the turbulent velocity dispersion.

\noindent Inserting into equation (A17) the analogue to equation (A14) leads
to the dispersion relation of axisymmetric perturbations of a gas disc,

\begin{equation}
1 = \frac{\lambda_{\rm crit}}{\mid\lambda\mid}\,\,\,
\frac{1}{1-({\frac{\omega}{\kappa})^2} +
{\frac{Q^2}{4}}\,({\frac{\lambda_{crit}}
{\lambda})^2}}, \qquad {\mbox{with}}\quad Q = \frac{c\kappa}{\pi G\Sigma_0}\,.
\end{equation}
The effect of the finite
thickness
of the disc can be taken into account analogous to equation (A15) by
multiplication with the reduction factor (A16).

\subsection{Stellar and gas disc combined}
In the case of a two--component disc the density perturbation in equation
(\ref{a14})
refers to the total surface density of the disc response. Thus, the density
response of the gaseous disc (A17) has to be added to the density response of the
stellar disc (A12) and the sum then inserted into equation (A14). The
resulting dispersion relation has the form

\begin{eqnarray}
1 &=& \frac{\mid\lambda\mid}{\lambda_{crit*}}\frac{3}{\xi_*}\,
\left\{1 - {{\omega / \kappa}\over {2\sin{(\pi \omega / \kappa)}}}
\int\limits_{-\pi}^{+\pi}dw '_1\,
{\exp (i {\omega \over\kappa} w'_1)\over \sqrt{1+{\frac{4}{3}}
\xi_* (\frac{\lambda_{\rm crit*}}{\lambda})^2 \cos^2({\omega '_1\over 2})}}
\right\}
 + \frac{\lambda_{crit,g}}{\mid\lambda\mid}\,\,\,
\frac{1}{1 -({\omega \over \kappa})^2 + \frac{Q_g^2}{4}\,({\frac{\lambda_{crit,g}}{\lambda})}^2}\,\,.
\end{eqnarray}

where the parameters ${\lambda_{crit*}}$, ${\xi_*}$ and $\lambda_{crit,g}$,
$Q_g$ are defined as in sections A1 and A2 for the stellar or the gaseous
discs, respectively. Setting $\omega = 0$ gives the curve of neutral stable
perturbations.

\end{document}